\documentclass[twocolumn,aps,superscriptaddress,nofootinbib]{revtex4-1}
\usepackage[T1]{fontenc}
\usepackage[latin9]{inputenc}
\usepackage{color}
\usepackage{array}
\usepackage{rotfloat}
\usepackage{units}
\usepackage{amsmath}
\usepackage{amssymb}
\usepackage{graphicx}
\usepackage{esint}
\usepackage{relsize}
\usepackage[unicode=true,pdfusetitle,
 bookmarks=true,bookmarksnumbered=false,bookmarksopen=false,
 breaklinks=false,pdfborder={0 0 1},backref=false,colorlinks=true]
 {hyperref}
\hypersetup{
  citecolor=blue}
\makeatletter
\usepackage{placeins}

\def\q2{q^2}
\def\Q2{Q^2}

\def\axion{a}
\def\gagg{g_{\axion \gamma \gamma}}
\def\gaee{g_{\axion e e}}

\def\ma{m_{\axion}}
\def\me{m_{e}}

\def\Ea{E_{\axion}}
\def\pa{p_{\axion}}

\def\Ga2g{\Gamma_{\axion \rightarrow \gamma \gamma}}

\newcommand\bea{\begin{eqnarray}}
\newcommand\eea{\end{eqnarray}}

\usepackage{array}
\usepackage{units}
\usepackage{bm}
\usepackage{braket}
\usepackage{slashed}
\usepackage[bbgreekl]{mathbbol}
\usepackage{bbm}

\makeatother

\begin{document}

\title{Constraining Axion-Like-Particles with germanium detector at the Kuo-Sheng Reactor Neutrino Laboratory}

\newcommand{\as}{Institute of Physics, Academia Sinica, Taipei 11529} 

\newcommand{\cusb}{Department of Physics,
Central University of South Bihar, Gaya 824236} 
\newcommand{\deu}{Department of Physics,
  Dokuz Eyl\"{u}l University, Buca, \.{I}zmir TR35160} 
\newcommand{\gla}{Department of Physics, Institute of Applied Sciences and Humanities, GLA University, Mathura 281406} 
\newcommand{\thu}{Department of Engineering Physics, Tsinghua University, 
Beijing 100084 }
\newcommand{\garw}{Department of Physics, H.N.B. Garhwal University, Srinagar 246174} 
\newcommand{\scu}{College of Physics, Sichuan University, Chengdu 610065} 
\newcommand{\NDHU}{Department of Physics, National Dong Hwa University, Shoufeng, Hualien
97401} 
\newcommand{\NTU}{Department of Physics, Center for Theoretical Physics, and Leung Center
for Cosmology and Particle Astrophysics, National Taiwan University,
Taipei 10617} 
\newcommand{\PDNTU}{Physics Division, National Center for Theoretical Sciences, National Taiwan University, Taipei 10617} 
\newcommand{\Tamkang}{Department of Physics, Tamkang University, New Taipei City 25137} 

\newcommand{\corrgc}{gchandrabhanu@gate.sinica.edu.tw}
\newcommand{\corrls}{lakhwinder@cusb.ac.in}
\newcommand{\corrhw}{htwong@phys.sinica.edu.tw}

\author{ Greeshma~C.}  \altaffiliation{\corrgc} \affiliation{\as}\affiliation{\cusb}  
\author{ L.~Singh }  \altaffiliation{\corrls}\affiliation {\cusb}
\author{ H.~T.~Wong } \altaffiliation{\corrhw } \affiliation{\as}
\author{ J.-W.~Chen}\affiliation{\NTU} \affiliation{\PDNTU}
\author{ M.~Deniz } \affiliation{\deu}
\author{ H.~B.~Li }  \affiliation{\as}
\author{ S. Karada\u{g}} \affiliation{\Tamkang}
\author{ S.~Karmakar } \affiliation{\as} \affiliation{\gla}
\author{ J.~Li }  \affiliation{ \thu }
\author{ F.~K.~Lin}   \affiliation{\as}
\author{ S.~T.~Lin } \affiliation{\scu}
\author{ C.-P.~Liu} \affiliation{\NDHU} \affiliation{\NTU}
\author{ S.~K.~Liu } \affiliation{\scu}
\author{ H.~Ma }  \affiliation{ \thu }
\author{ D.~K.~Mishra}  \affiliation{\as}\affiliation{\cusb}
\author{ M.~K.~Pandey} \affiliation{\NTU} 
\author{ K.~Rani} \affiliation{\cusb}
\author{ R.~Raj} \affiliation{\cusb}
\author{ V.~Sharma}\affiliation{\garw}
\author{ M.~K.~Singh } \affiliation{\as} 
\author{ M.~K.~Singh } \affiliation{\gla}
\author{ V.~Singh }  \affiliation{\cusb}
\author{ C.-P.~Wu}\affiliation{\NDHU}
\author{ L.~T.~Yang } \affiliation{ \thu }
\author{ Q.~Yue } \affiliation{ \thu }

\collaboration{TEXONO Collaboration}


\begin{abstract}
  We present results of a search for Axion-Like Particles (ALPs) produced $via$
  Primakoff and Compton-like scattering channels, using data acquired with  
  TEXONO experiment at the Kuo-Sheng Nuclear Power Station. The analysis is based on
  278.91 days of reactor-ON and 43.60 days of reactor-OFF data. These datasets were
  collected using a 1.06 kg high-purity germanium detector located 28 m from a 2.9 GW reactor core.
  No significant excess is observed in the residual spectrum from the reactor-ON and reactor-OFF data subtraction.
  Using data acquired with low-background germanium detectors, upper limits on both the ALP-photon
  ($\gagg$) and ALP-electron ($g_{aee}$) couplings are derived for ALP masses ranging from 1 eV to 3 MeV at 90\% confidence level.
  Since both $\gagg$ and $g_{aee}$ couplings contribute to ALP production and detection, a combined analysis is performed
  by treating both channels as active parameters.

\end{abstract}

\maketitle
\section{Introduction}
The quantum chromodynamics (QCD) axion is proposed by Peccei and Quinn in $1977$ to solve the strong-CP problem~\cite{Peccei:1977hh,Wilczek:1977pj,Weinberg:1977ma}. The mass and coupling of the QCD axion are intrinsically related through the Peccei-Quinn symmetry-breaking scale. On the other hand, Axion-Like Particles (ALPs)
constitute a broader class of
pseudo-Nambu-Goldstone bosons that arise in various extensions of the
Standard Model (SM), particularly in string theory compactifications~\cite{Dine:1986zy,Becker:1995kb}. Unlike the QCD axion, ALPs are not required to solve the strong-CP problem, and
their mass and coupling are independent parameters. 

ALPs are also promising candidates for cold dark matter~\cite{CDM}. Searches for ALPs are intensely
pursued over a broad range of experimental parameter space and astrophysical environments~\cite{RAFFELT1982323,Cameron:1993mr,Graham:2015ouw,  Irastorza:2018dyq,Sikivie:2020zpn,Choi:2020rgn,ParticleDataGroup:2022pth,AxionLimits}.

Techniques adopted for ALP experiments are diverse and complementary~\cite{Choi:2020rgn,ParticleDataGroup:2022pth, Graham:2015ouw}. Searches for dark matter ALPs are conducted using microwave cavity~\cite{PRL:ADMX, PRL:Braine, PRL:MADMAX, PRD:Masha} and as part of direct detection~\cite{XENON:2024znc, SuperCDMS:2019jxx, PandaX:2024sds} experiments.

Solar ALPs with masses up to 20 keV are searched with the helioscope experiments~\cite{CAST:2008ixs, CAST:2017uph, Minowa:1998sj}.
The production of ALPs with intense photon sources inspires ``light-shining-through-walls'' experiments~\cite{ParticleDataGroup:2022pth, Bahre:2013ywa}, and interferometry techniques~\cite{ParticleDataGroup:2022pth,Liu:2018icu, Oshima:2023csb}, which are sensitive to the sub-eV mass regime.
ALPs can also be probed indirectly through gamma-ray observatories~\cite{ParticleDataGroup:2022pth, Schiavone:2025dff, Alfaro:2026hca, Egorov:2020cmx, Long:2021udi, Budnev:2022ksa, Huang:2015fca}.
ALPs in the GeV-scale mass range can be probed by beam-dump and fixed-target
experiments~\cite{Dobrich:2015jyk, Feng:2018pew, Berlin:2018bsc, Volpe:2019nzt,Berlin:2018pwi, Alekhin:2015byh}. 

Among SM particles, ALPs are generally expected to couple to photons and electrons. These interactions are foundational
to their phenomenology, providing the primary channels for their production in extreme environments; therefore, 
nuclear reactors may provide intense sources of ALPs in the keV to MeV mass range.
Within the nuclear reactor core, ALPs may be produced through nuclear de-excitation, Compton-like
scattering, and Primakoff conversion~\cite{JHEP_2021}. Due to their weak interactions with SM particles,
ALPs can propagate from the reactor core to nearby detectors with negligible attenuation.
This allows sensitive probes of their couplings over a broad region of the parameter space.

In our earlier study of reactor-based axions~\cite{TEXONO:2006spf}, we considered the production of axions through neutron capture
and nuclear de-excitation, in competition with the standard photons emission. In our recent work~\cite{OurPRDALP},
we identified novel ALP detection channels $via$ inelastic inverse Primakoff processes associated with atomic ionization and excitation.
The cross sections for these novel channels were evaluated using benchmark atomic many-body methods for non-relativistic
dark matter ALPs. We found that the inelastic inverse Primakoff ionization channel provides the dominant contribution
for $\Ea<$~1~keV.  
In the present work, we adopt the studies of Ref.~\cite{NEON:2024kwv, MINER:2016igy, Arias-Aragon:2023ehh} and extend to reactor ALPs which are relativistic and produced by both Primakoff and Compton-like processes. 

This report is organized as follows: Section~\ref{sec:formulation} formulates the interaction mechanisms and presents the expected production and detection rates in a high-purity germanium (HPGe) detector located at the Kuo-Sheng Reactor Neutrino Laboratory (KSNL)~\cite{TEXONO:2018}.
Section~\ref{sec:Setup} summarizes the experimental setup and data acquisition system. Section~\ref{sec:Result} presents our results in comparison with existing experimental and astrophysical constraints. Our results are obtained by independently evaluating effects induced by $\gagg$ and $\gaee$ couplings, as well as by performing a combined analysis in which both couplings are simultaneously active.

\begin{figure}[h!]
  {\bf (a)}\\
  \vspace{0.2cm}
  \includegraphics[width=4.2cm]{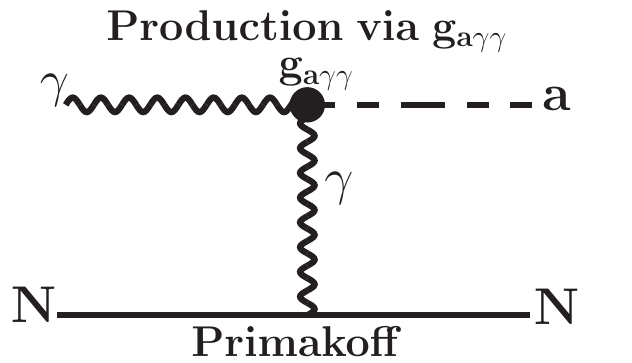}
  \includegraphics[width=4.3cm]{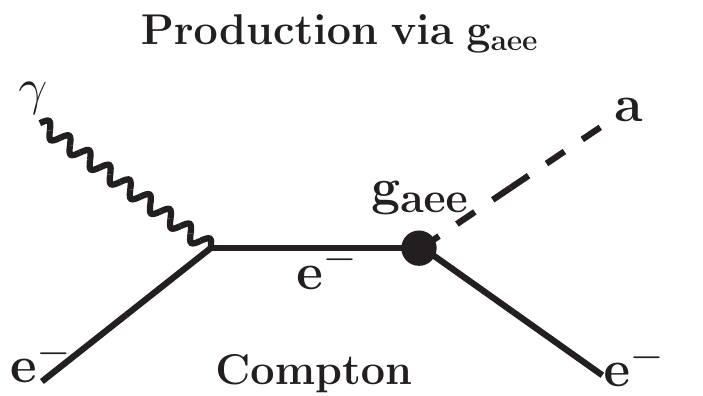}\\
  \vspace{0.8cm}  
  {\bf (b)}\\
  \vspace{0.2cm}
  \includegraphics[width=4.3cm]{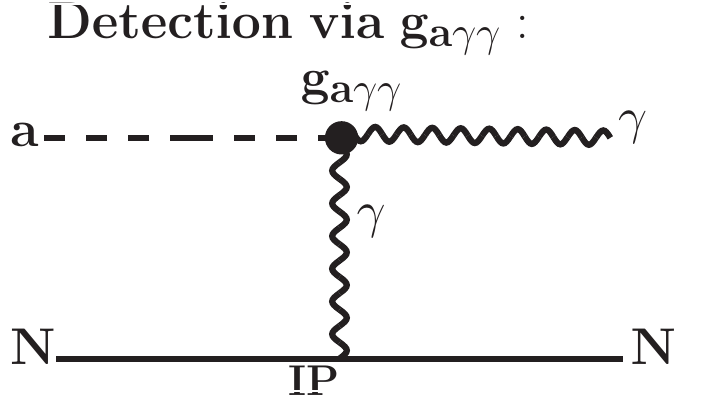}
  \includegraphics[width=4.2cm]{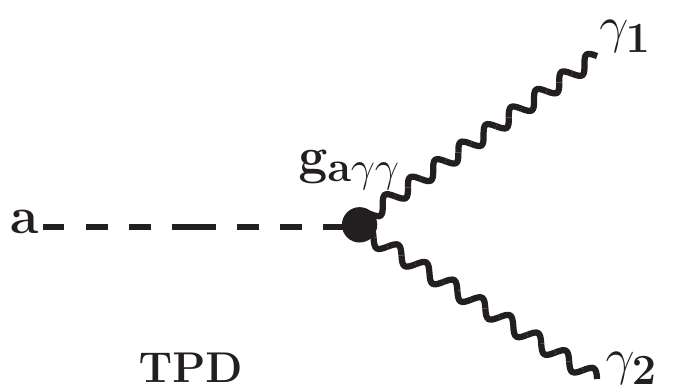}\\
  \vspace{0.8cm}  
  {\bf (c)}\\
  \vspace{0.2cm}
  \includegraphics[width=5.3cm]{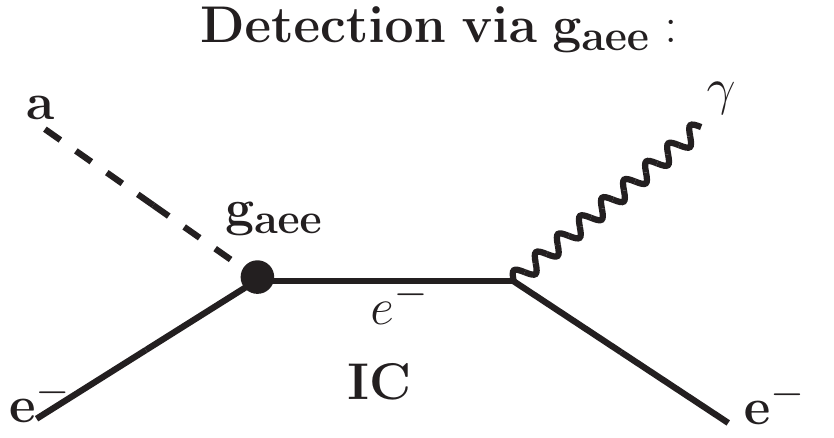}
  \includegraphics[width=5.3cm]{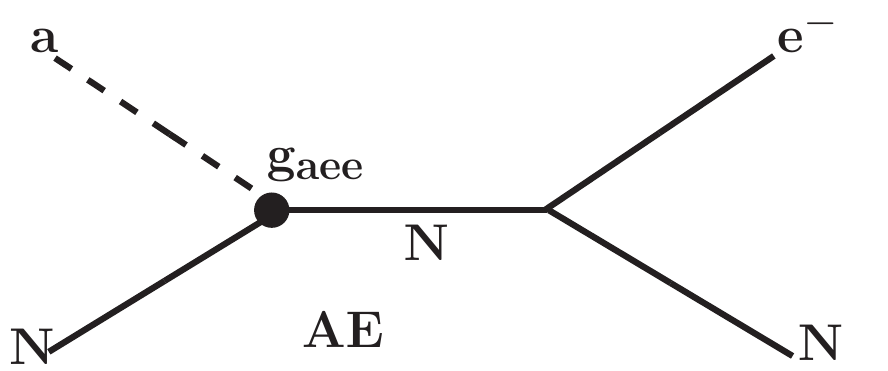}
  \vspace{0.8cm}  
  \includegraphics[width=5cm]{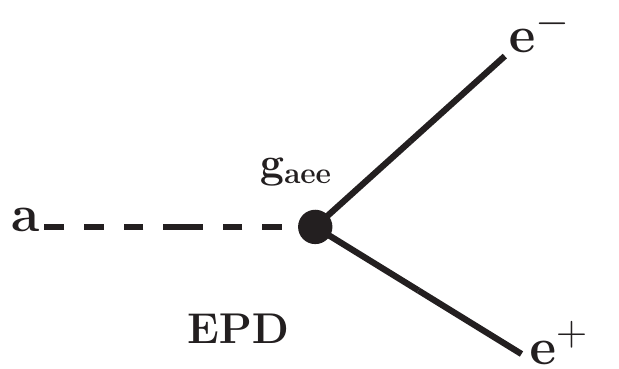}
  \caption{Feynman diagrams showing (a) The production mechanisms of reactor ALPs: Primakoff and Compton
    (b) The detection mechanisms of reactor ALPs via $\gagg$: inverse Primakoff (IP) and two photon decay (TPD); and (c)
    The detection mechanisms of reactor ALPs via $\gaee$: inverse Compton (IC), axio-electric absorption (AE),
    and electron-positron decay (EPD). } 
\label{fig:Feynman}
\end{figure}

\section{Formulation}
\label{sec:formulation}
The low-energy interaction Lagrangian density describing the coupling of the ALP
field ($a$) to the photon field ($A_{\mu}$) and SM fermion fields
($\psi_{f}$) is typically expressed as follows~\cite{ParticleDataGroup:2024cfk}:
\begin{equation}
\mathcal{L}_{I}=-\frac{\gagg}{4} a F_{\mu\nu}\widetilde{F}^{\mu\nu} - {\mathlarger{\sum}_{f}} \frac{g_{aff}}{2m_{f}} (\partial_{\mu} a) \bar\psi_f \gamma^{\mu} \gamma_5\psi_f,\label{eq:L_agg}
\end{equation}
where $F^{\mu\nu} {=} \partial^{\mu}A^{\nu} {-} \partial^{\nu}A^{\mu}$ and
$\widetilde{F}^{\mu\nu} {\equiv} \frac{1}{2}\epsilon^{\mu\nu\rho\sigma}F_{\rho\sigma}$
are the photon field strength tensor and its dual, respectively, with $\epsilon^{0123} = 1$.
The parameter $\gagg$ denotes the strength of the ALP-photon coupling in units of $\rm GeV^{-1}$, while
$g_{aff}$ denotes the dimensionless ALP-fermion coupling.

The ALP production and interaction channels described by Eq.~\ref{eq:L_agg} and relevant to this 
analysis are depicted in Figure~\ref{fig:Feynman}.
\begin{itemize}
\item
  The $\gagg$ coupling induces ALP production through
  Primakoff scattering, 
\begin{equation}
\gamma + A \rightarrow a + A,  \label{eq:prim}  
\end{equation}
where $A$ represents an atomic
target. This coupling also enables detection through the corresponding inverse Primakoff process, 
as well as ALP decay into photon pairs.
\item
The $\gaee$ coupling involves a
more diverse phenomenology. Production primarily occurs through Compton-like 
scattering,
\begin{equation}
\gamma + e^{-} \rightarrow a + e^{-}, \label{eq:comp}      
\end{equation}
while detection can be proceed via either the inverse Compton-like process or axio-electric absorption (analogous to the photoelectric 
effect).
In kinematically allowed regions, this coupling also permits ALP decay into 
$e^{-}e^{+}$ pairs.
\end{itemize}

Each of these channels is associated with distinct signatures,
allowing reactor-based experiments to probe complementary ALP parameter space 
compared with other experimental approaches.

\subsection{Reactor-ALP Flux}
Photons are produced abundantly within reactor cores. Approximately 50\% of the $\gamma$-ray activity originates
from $"prompt-emissions"$, generated directly by highly excited fission fragments. The remaining contribution
arises from secondary processes, including the radiative de-excitation of daughter nuclei, inelastic neutron
scattering, and neutron capture reactions in core structural materials~\cite{Altmann:1995bw}. The prompt
$\gamma$-ray spectrum of a reactor core, expressed in units of $\text{MeV}^{-1} \text{s}^{-1}$,
is characterized
by the following distribution for $E_{\gamma}$ $\ge$ 200 keV~\cite{Bechteler1984},
\begin{equation}
  \frac{d\phi_{\gamma}}{dE_{\gamma}} = 0.58 \times 10^{21} e^{-1.1 \left(E_{\gamma}(\rm{MeV})\right)} \times \text{Power (GW)}.
  \label{Eq:photon_flux}
\end{equation}
This spectral model has been adopted in earlier studies of reactor axions~\cite{Altmann:1995bw},
ALPs~\cite{NEON:2024kwv}, dark photons~\cite{Ge:2017mcq, Park:2017prx}, and reactor millicharged
particles~\cite{TEXONO:2018nir}. The abundant production of photons naturally leads to their interactions
with reactor core materials and atomic electrons. The primary constituent of
reactor fuel is $^{238}$U, which undergoes gradual and continuous depletion as a function of
both operational time and fuel burnup. The photon interactions that contribute to ALP production are illustrated in Fig.~\ref{fig:Feynman}(a).

\subsubsection{Production by ALP-Photon Coupling}

The differential cross section for the Primakoff production of ALPs from a
nucleus ($N$) is expressed as~\cite{PhysRevLett.123.071801}, 
\begin{multline}
\frac{d\sigma^{P}}{dt} = \frac{2\alpha Z^2 |F(t)|^2 \gagg^2 M_N^4}{t^2 (M_N^2 - s)^2 (t - 4M_N^2)^2}  \times  \\ 
\bigg[\ma^2t (M_N^2 + s) - \ma^4 M_N^2 - t\left((M_N^2 - s)^2 + st\right)\bigg],
\label{Eq::dcsPrimakoff}
\end{multline}
where $t$ and $s$ represent the Mandelstam variables, $\ma$ is the ALP mass, and $Z$ and $M_{N}$ denote the atomic
number and nuclear mass, respectively. The symbol $\alpha$ represents the fine-structure
constant, and $|F(t)|$ is the form factor. In the low-momentum-transfer region, 
$|q|$~=~$\sqrt{-t}$~$\le$~$\sqrt{7.39 m_{e}^{2}}$, where $m_{e}$ is the electron mass, we adopt the Thomas-Fermi-Moli\`ere atomic form
factor~\cite{Moliere1947}, which is well approximated by the function: 
\begin{equation}
  F(q)  =  Z \sum_{i = 1}^{3} \alpha_i \left[1 + \left(\frac{q a_{TF}} {\beta_{i}}  \right)^{2}\right]^{-1},
  \label{eq:apprx}
\end{equation}
where $a_{TF} = a_{0}(9\pi^{2}/128)^{1/3} Z^{-1/3}$ is the Thomas-Fermi screening length, and $a_0$ is the Bohr radius.
The Moli\`ere coefficients are dimensionless and are given by $\alpha_{1} = 0.35$, $\alpha_{2} = 0.55$, $\alpha_{3} = 0.10$, 
$\beta_{1} = 0.3$, $\beta_{2} = 1.2$, and $\beta_{3} = 6.0$. At large momentum transfers ($|q|>\sqrt{7.39 m_{e}^{2}}$),
atomic screening becomes negligible, and the nuclear Helm form factor is applied.

\begin{figure}[t]
\includegraphics[width=8cm]{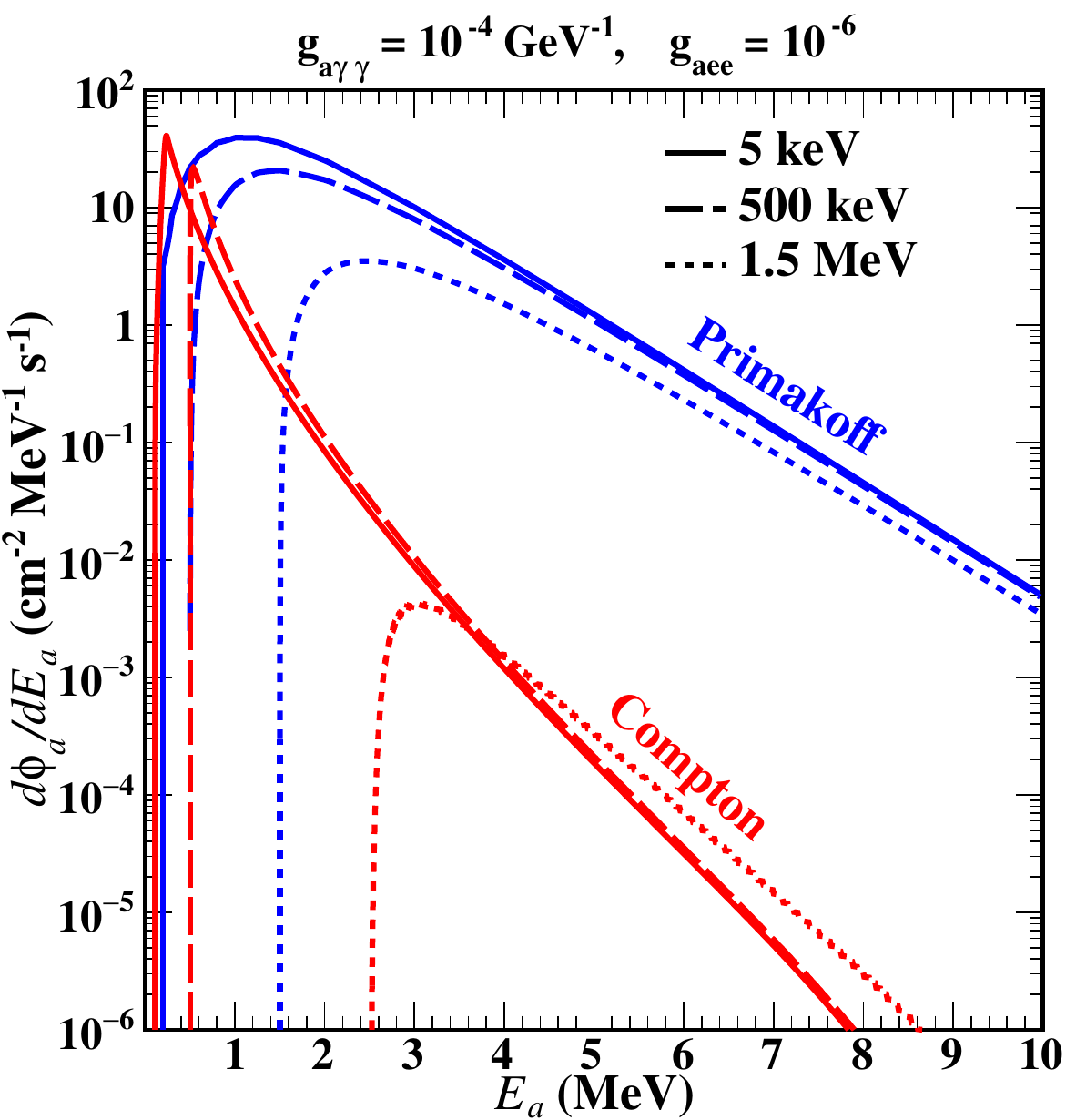}
\caption{Comparison of the differential reactor ALP fluxes from the Primakoff and Compton processes at a distance of 28 m from a 2.9 GW reactor core, shown for 
$m_{a}$ = 5 keV, 500 keV and 1.5 MeV.}
\label{fig:flux}
\end{figure}

The photons generated in a nuclear reactor typically have energies
up to 10 MeV, which are several orders of magnitude smaller than the mass of a uranium nucleus.
Therefore, the nuclear recoil resulting from photon interactions is negligible, allowing the total energy
of the produced ALPs to be approximated as $\Ea$ = $E_{\gamma}$. 
Analytically, this simplification enables us to derive the ALP flux directly from the photon distribution
defined in Eq.~\ref{Eq:photon_flux} as
\begin{equation}
\frac{{d\phi}^{P}_a}{d\Ea} =  \frac{1}{4 \pi R^{2}} \left(\frac{\sigma_{\rm Prim}}{\sigma_{\rm tot}}\right) \frac{{d\phi}_{\gamma}(\Ea)}{d\Ea}, 
   \end{equation}
where $R$ is the distance from the center of reactor core to the detector. The total cross section, $\sigma_{\rm tot}$,
is the sum of SM cross section ($\sigma_{\rm SM}$),
and the total Primakoff scattering cross section,
($\sigma_{\rm Prim}$). The latter cross section is obtained by integrating Eq.~\ref{Eq::dcsPrimakoff} over the allowed kinematic range.
The $\sigma_{\rm SM}$ value is adopted from the photon cross-section
database~\cite{PhotonCS}. The ALP flux produced via the Primakoff process at a distance of 28 m from a 2.9 GW reactor core is presented in Fig.~\ref{fig:flux} for different ALP masses, assuming $g_{a\gamma\gamma} = 10^{-4}\text{ GeV}^{-1}$.

\subsubsection{Production by ALP-Electron Coupling}

When $\gaee$ is non-zero, ALPs may be produced through the
Compton-like process as described in Eq.~\ref{eq:comp}, in which $\gamma$-rays
with $\mathcal{O}$(MeV) energies scatter off electrons in the reactor core.
The differential production cross section of the Compton-like process is given by \cite{Brodsky:1986mi},
\begin{multline}
\frac{d\sigma^{C}}{dE_a}
=
\frac{Z\, \pi\, g_{aee}^2 \alpha\, x}
     {4\pi (s-m_e^2)(1-x) E_{\gamma_{in}}}
\\[4pt]
\times
\left[
x
- \frac{2 m_a^2 s}{(s-m_e^2)^2}
+ \frac{2 m_a^2}{(s-m_e^2)^2}
  \left(
     \frac{m_e^2}{1-x}
     + \frac{m_a^2}{x}
  \right)
\right],
\label{Eq::ComptProd}
\end{multline}
where $s =  m^{2}_{e} +  2 m_{e} E_{\gamma_{in}}$ is the usual Mandelstam
variable, and $E_{\gamma_{in}}$ is the energy of the incoming photon, and
\begin{equation}
    x = 1 - \frac{\Ea}{E_{\gamma_{in}}} +\frac{\ma^2}{2 E_{\gamma_{in}} \me}.
\end{equation}
Unlike the Primakoff process, Compton scattering is an inelastic process; therefore,
$\Ea \neq E_{\gamma_{in}}$. The energy threshold for an incoming photon to produce an ALP
of mass $m_a$ is given by
\begin{equation}
    E_{\gamma_{in}}^{min} = \frac{2 \me \ma + \ma^2}{2 \me}. 
\end{equation}

The differential ALP flux produced via the Compton-like process is evaluated by convolving
the reactor $\gamma$-ray spectrum with the differential production cross section given in Eq.~\ref{Eq::ComptProd},
normalized by the total cross section ($\sigma_{\rm tot}$),
\begin{equation}
\frac{{d\phi}^{C}_a}{d\Ea} = \frac{1}{4 \pi R^{2}} \int_{E^{min}_{\gamma_{in}}}^{E^{max}_{\gamma_{in}}} \frac{1}{\sigma_{\rm tot}} \frac{d\sigma^{C}}{dE_a}\frac{{d\phi}_{\gamma}}{dE_{\gamma_{in}}} dE_{\gamma_{in}}.
\end{equation}

The differential energy spectra of reactor ALP fluxes produced through the Compton-like processes for different masses are presented in Fig.~\ref{fig:flux}. The fluxes are evaluated at a distance of 28 m from the 2.9 GW nuclear reactor core, assuming coupling constants of $\gaee = 10^{-6}$.

\begin{figure}[t]
\includegraphics[width=8cm]{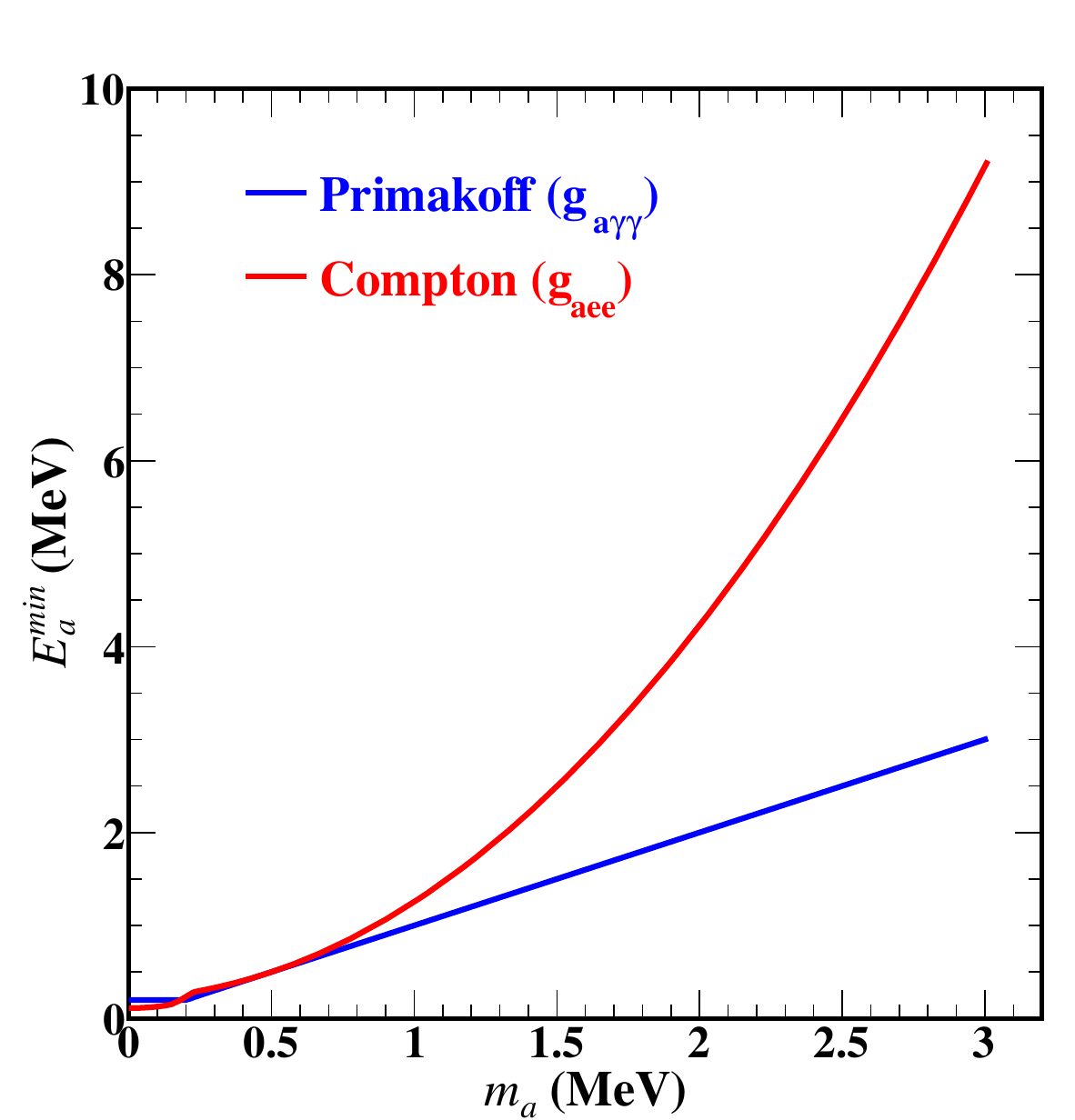}
\caption{Comparison of ${E_{a}}^{min}$ values for Primakoff and Compton production processes for ALP masses ranging from 1 eV to 3 MeV.}
\label{fig:Eamin}
\end{figure}

The minimum energy, ${E_{a}}^{min}$, of an ALP produced through the Primakoff or Compton-like processes is determined by
four-momentum conservation and the kinematics of the final state. In the Primakoff process, an incident photon is converted
into an ALP in the static Coulomb field of a heavy nucleus. Owing to its large mass ($M \gg E_\gamma$), the nucleus absorbs
the required momentum with negligible recoil. Therefore, in the forward-scattering limit, the minimum ALP energy is approximately equal to its rest mass ($E_{a}^{min} \approx m_a$). In the Compton-like process, however, electron recoil is significant during the interaction. In the laboratory frame, the minimum energy of the emitted ALP is given by
\begin{align}
  {E_{a}}^{min} & = \frac{1}{m_{e}^{2} + 2m_{e}E_{\gamma} }\left[(m_{e} + E_{\gamma})(m_{e}E_{\gamma} + \frac{m_{a}^{2}}{2}) \right.\nonumber  \\ 
    & \left. - E_{\gamma}\sqrt{ (m_{e}E_{\gamma} + \frac{m_{a}^{2}}{2})^{2} - (m_{e} + E_{\gamma})^{2}m_{a}^{2} + E_{\gamma}^{2}m_{a}^{2}}\right], 
\end{align}
where $E_{\gamma}$ is the incoming photon energy, which must be greater than $m_{a}$. The variation of ${E_{a}}^{min}$ for ALP masses ranging from 1 eV to 3 MeV is shown in Fig.~\ref{fig:Eamin}.

\subsection{Detection of Reactor ALPs}
The detection channels of ALPs depend on their couplings to photons, electrons, and nucleons. Direct detection through nucleon coupling, $g_{aNN}$, relies exclusively on nuclear excitation, which requires the emitted ALP to be resonantly-absorbed by the same species of daughter nucleus from which it was produced. Because constructing a detector from this specific daughter isotope is experimentally not feasible, therefore, the detection must proceed through either $g_{aee}$- or $g_{a\gamma\gamma}$-induced processes. As a result, any constraints on $g_{aNN}$ would be expressed in terms of products of paired couplings, namely $|g_{aee} \times g_{aNN}|$ or $|g_{a\gamma\gamma} \times g_{aNN}|$. This parameter space was thoroughly investigated in our previous work~\cite{TEXONO:2006spf}. Therefore, we exclude the $g_{aNN}$ coupling from the present analysis.

\subsubsection{Detection via ALP-Photon Coupling}
When ALPs possess a non-zero photon coupling $\gagg$, the inverse Primakoff (IP) scattering 
and spontaneous two-photons decay (TPD) are two primary detection channels, as illustrated in Figure~\ref{fig:Feynman}(b).
\begin{enumerate}
\item
  {\bf IP: $a + A \rightarrow \gamma + A$}. \\

The total cross section for IP scattering is adopted from Ref.~\cite{Abe:2020mcs} and is given as,
\begin{align}
\sigma_{IP}
&=
\frac{\alpha\, \gagg^{2}}{8}
\int_{\Ea - \pa}^{\Ea + \pa} \! dq \,
\frac{\bigl[q^{2} - (\Ea - \pa)^{2}\bigr]}{\pa^{2} q^{3}}
\nonumber \\[6pt]
&\quad\times
\bigl[(\Ea + \pa)^{2} - q^{2}\bigr] \bigl|\, Z - F(q) \,\bigr|^{2}.
\label{eq:sigma_el}
\end{align}
For momentum transfers up to $q \le 32$~keV, relativistic Hartree-Fock (RHF) atomic form factors for all elements are tabulated in
Ref.~\cite{formfactor}. Since the Thomas-Fermi-Moli\`ere form factor provides good agreement with the RHF approximation and also remains valid for
$q \ge 32$~keV range, we adopt it for germanium to calculate the scattering cross section.

\item
{\bf TPD: $a \rightarrow \gamma \gamma$ } \\

ALPs may undergo two-photon decay with the well-known decay width,
\begin{equation}
  \Ga2g=\frac{1}{64\pi}\gagg^{2}\ma^{3}.
\end{equation}

The sensitivity of the detector is mainly governed by characteristic decay length of ALPs, $L_{a} = v_{a} \tau_{a}$,  where $v_{a} = |p_{a}|/\Ea$ is the relativistic velocity and $\tau_{a} = \Ga2g^{-1} $ is the ALP lifetime. This decay length should be
comparable to or greater than the baseline between the reactor core and detector, $R$. The survival probability ($\mathcal{P}_{\text surv}$) of ALPs can be expressed as
\begin{eqnarray}
  \mathcal{P}_{\text surv} = exp{\left(\frac{-R}{L_{a}} \right)}.
\end{eqnarray}
The total expected differential event rate from these two distinct channels is given by
\begin{align}
  \frac{dR_T}{d\Ea} & = \mathcal{P}_{\text surv} N_{T} \sigma_{IP} \frac{{d\phi}^{P}_a}{d\Ea} + \mathcal{P}_{\text surv} A \frac{{d\phi}^{P}_a}{d\Ea}\mathcal{P}_{\text decay}
  \label{Eq:gaggRate}
\end{align}
where $N_{T}$ and $A$ are the number of target nuclei and transverse area of the detector, respectively.
The probability that an ALP decays inside the detector, $\mathcal{P}_{\text decay}$, is given by
\begin{equation}
    \mathcal{P}_{\text decay} = 1 - exp{\left(\frac{-L_{det}}{L_{a}}\right)},
\end{equation}
where $L_{det}$ is the longitudinal length of the detector along the incident ALP beam direction.

\end{enumerate}
\begin{figure}[t]
  \includegraphics[width=9cm]{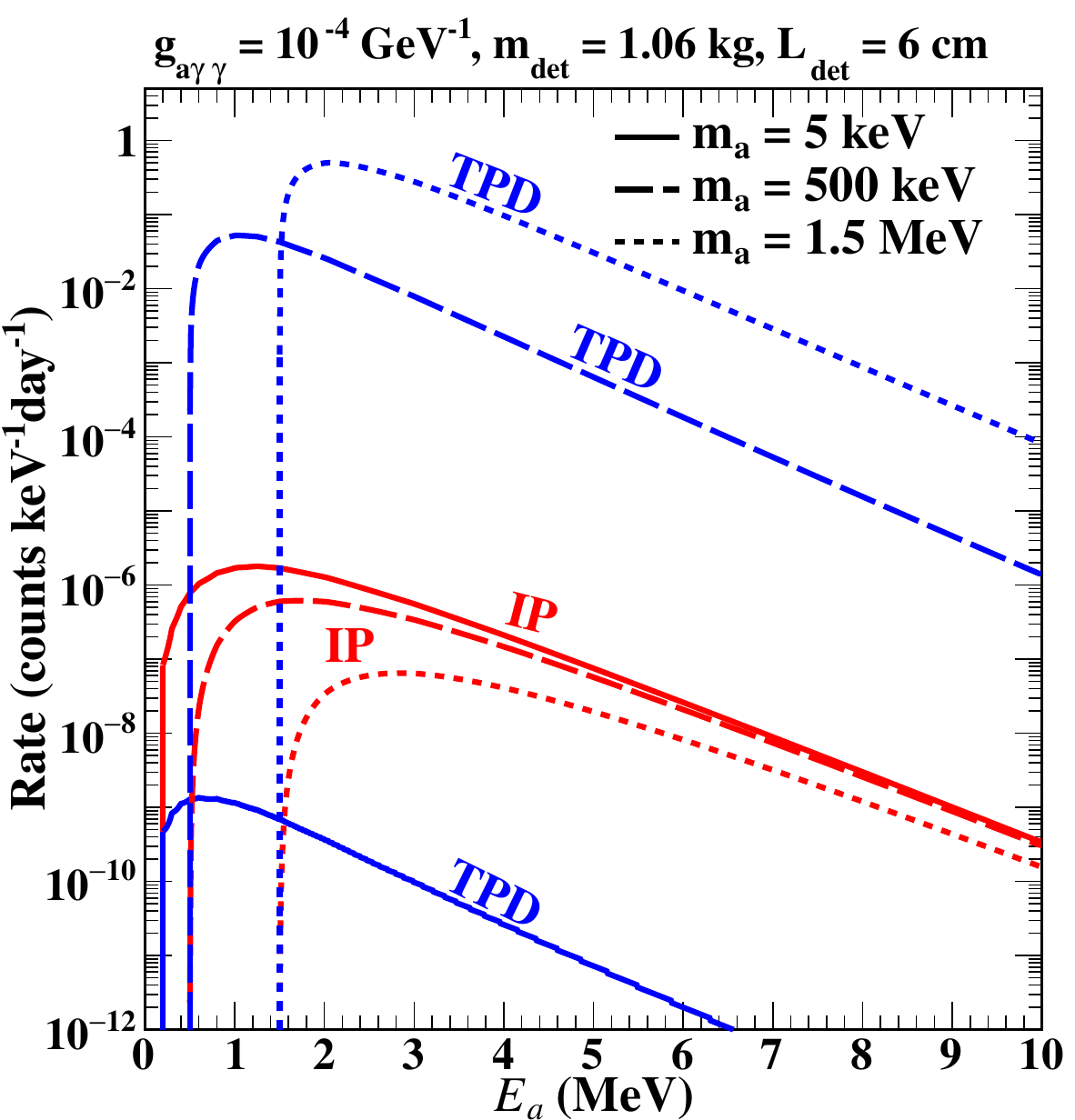}
  \caption{Differential event rates of reactor ALPs for the IP and TPD channels for a germanium
    target, taking $\gagg = \rm {10^{-4}~GeV^{-1}}$ and several representative values of $\ma$.}
  \label{fig:gaggRates}
\end{figure}
Figure~\ref{fig:gaggRates} compares the differential event rates of different photophilic ALP detection channels 
in a germanium detector. The rates are calculated under the identical reactor conditions, assuming $\gagg = 10^{-4}~\mathrm{GeV^{-1}}$, a detector mass of ${m_{det}} = 1.06$ kg, and a detector length of ${L_{det}} = 6$ cm. Three representative ALP mass values are selected to illustrate the dominance of specific detection channels.

 \begin{figure}[t]
   \includegraphics[width=9cm]{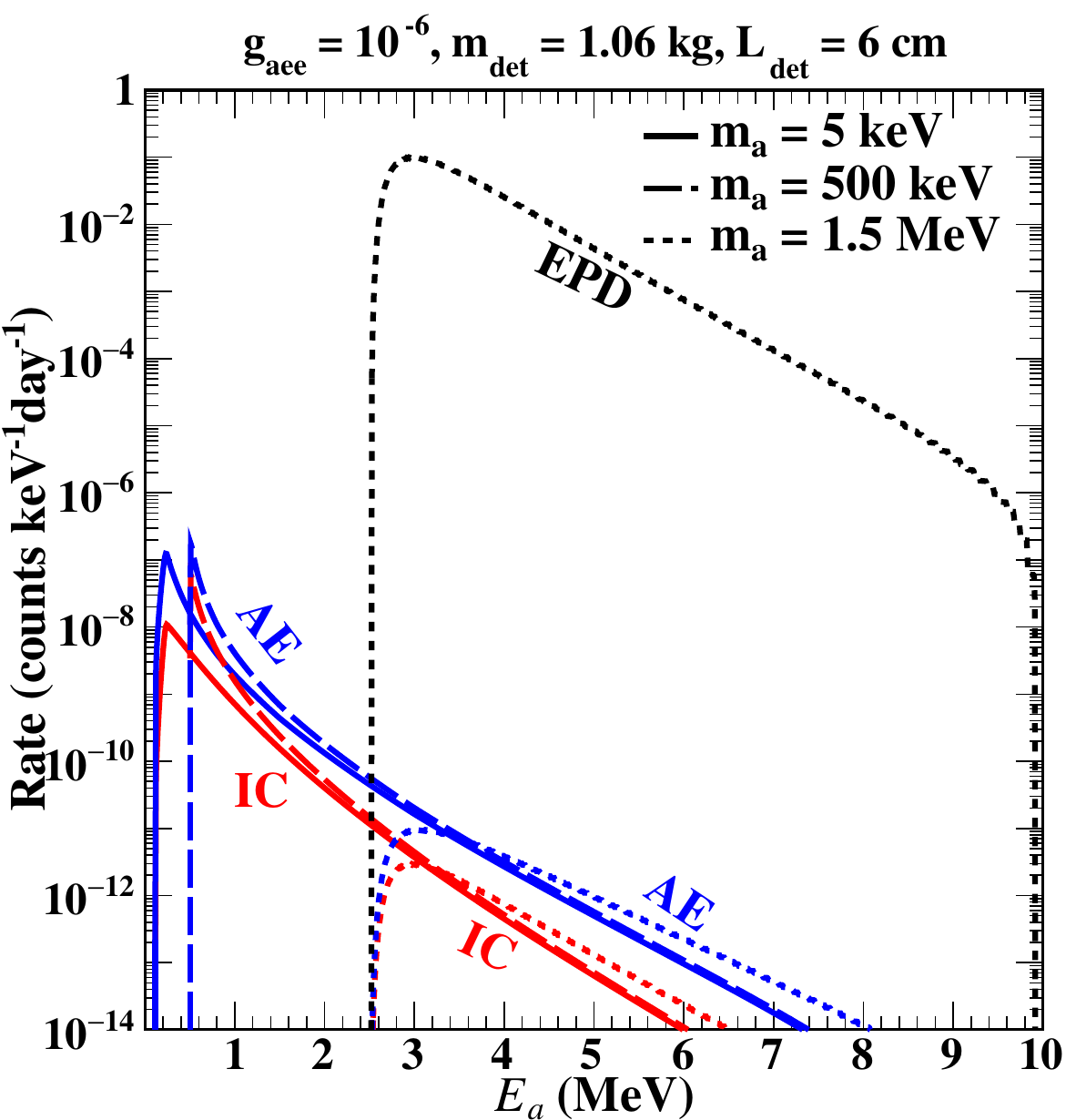}
   \caption{Differential event rates of reactor ALPs for the IC scattering, AE absorption and EPD for a germanium
     target, taking $\gaee = \rm {10^{-6}}$ and several representative values of $\ma$.}
   \label{fig:gaeeRates}
 \end{figure}

\subsubsection{Detection via ALP-Electron Coupling}

As depicted in Figure~\ref{fig:Feynman}(c), 
for $\gaee \ne 0$ and $\gagg$=0, the prominent ALP detection channels include
inverse Compton-like scattering (IC), Axio-electric absorption (AE) and decay into 
an electron-positron pair (EPD):

\begin{enumerate}
\item
  {\bf IC:  $a + e^- \rightarrow \gamma + e^-$}.\\

  The IC process produces an observable photon and a recoil electron and, both of which can
deposit their entire energy within the germanium detector. The total cross section for the IC process is expressed as~\cite{Avignone:1988bv}:
\begin{equation}
\begin{aligned}
\sigma_{IC} & = \frac{g_{aee}^2 \alpha}{8 m_{e}^2 \pa} \left[ \frac{2m_{e}^2(m_{e}+\Ea)y}{(m_{e}^2+y)^2} \right. \\
& + \frac{4m_{e}(\ma^4 + 2\ma^2 m_{e}^2 - 4m_{e}^2 \Ea^2)}{y(m_{e}^2+y)} \\
& \left. + \frac{4m_{e}^2 \pa^2 + \ma^4}{p_a y} \ln \frac{m_{e} + \Ea + \pa}{m_{e} + \Ea - \pa} \right],
\end{aligned}
\end{equation}
where $y~=~2m_{e} \Ea + m_{a}^{2}$.
\item
  {\bf AE:  $a + e^- + Z \rightarrow e^- + Z$} \\

  An ALP is absorbed by a bound electron, resulting in the emission of an electron with an energy of
  $\Ea - E_{b}$, where $E_{b}$ is the binding energy.
  The AE cross section
  ($\sigma_{AE}$) is related to the photoelectric cross section ($\sigma_{PE}$) as~\cite{Pospelov:2008jk}:
  \begin{equation}
    \sigma_{AE} = \sigma_{PE} \frac{g_{aee}^2}{16\pi\alpha m_e^2} \frac{3E_a^2}{\beta_a} \left( 1 - \frac{\beta_a^{2/3}}{3} \right),
  \end{equation}
  where $\beta_a = |\vec{p}_a|/E_a$ denotes the ALP velocity.
  The AE cross section ($\sigma_{AE}$) dominates at low energies because it scales directly with $\sigma_{PE}$.
  Although $\sigma_{PE}$ decreases rapidly with increasing  ALP energy, the $\Ea^{2}$ dependence in $\sigma_{AE}$  provides
  a compensating enhancement.

\item
{\bf  EPD:  $a \rightarrow e^- e^+$}\\

The EPD channel is kinematically accessible only when the ALP mass exceeds
the threshold $\ma > 2\me$. The AE channel remains dominant over IC scattering in germanium up to
$E_{a} =2m_{e}$. Once $\Ea$ exceeds this threshold, the EPD channel is kinematically allowed; and
this channel rapidly dominates over both the AE and IC channels. The decay width for the EPD process is given by
\begin{equation}
\Gamma_{a\rightarrow e^{-} e^{+}} = \frac{g_{aee}^{2} m_{a}}{8 \pi} \sqrt{1-\frac{4m_{e}^{2}}{\ma^{2}}}.
\end{equation}

The expected total differential signal rate at the detector for the $g_{aee}$ coupling can be expressed as
\begin{align}
\frac{dR_{T}}{d\Ea} & =  \mathcal{P}_{\text surv}N_{T} \sigma_{IC} \frac{{d\phi}^{C}_a}{d\Ea} + \mathcal{P}_{\text surv}A \frac{{d\phi}^{C}_a}{d\Ea}P_{decay} \nonumber \\
& + \mathcal{P}_{\text surv}N_{T} \sigma_{AE} \frac{{d\phi}^{C}_a}{d\Ea}
\label{Eq:gaeeRate}
\end{align}

\end{enumerate}

The differential event rates for the IC, AE and EPD channels corresponding to three representative ALP masses are
shown in Fig.~\ref{fig:gaeeRates}. In the kinematically allowed region, the EPD channel dominates significantly over the other two channels.

\section{Experimental Constraints}
\label{sec:ExptSetup}

This analysis is based on high-energy data collected with the HPGe detector~\cite{TEXONO:2006xds}
at KSNL as a part of the TEXONO program. Experimental details are described in  Refs.~\cite{TEXONO:2006xds, TEXONO:2009knm, TEXONO:2018}, and the references therein.

\begin{figure}[!ht]
\includegraphics[width=9cm]{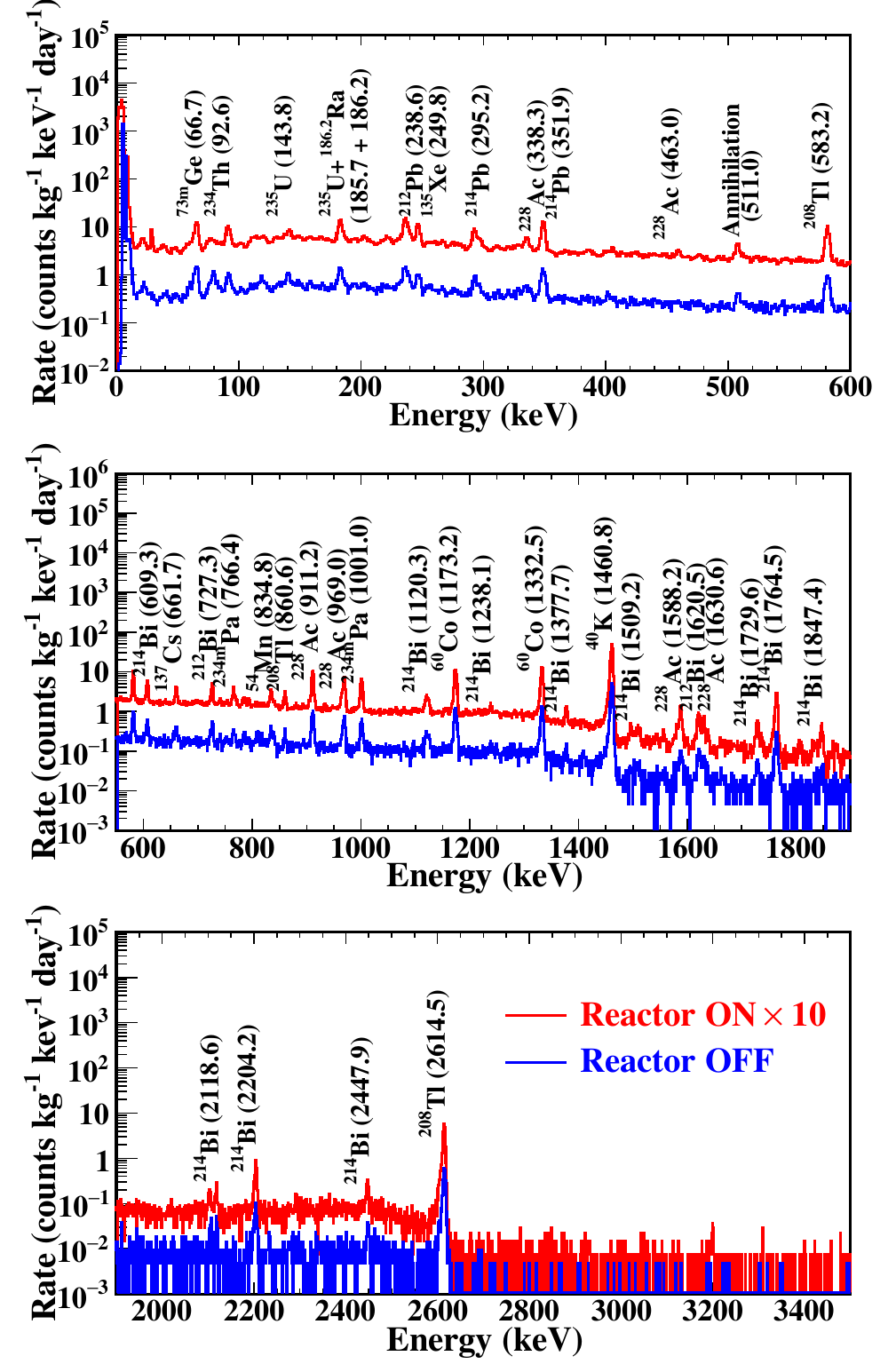}
\caption{Measured $AC^{-}~\otimes~CR^{-}$ selection spectra for Period-III over the full analysis energy range. All
  prominent background $\gamma$-ray lines are identified. The reactor-ON and reactor-OFF data are presented separately to
  facilitate comparison.}
\label{fig:DataHist}
\end{figure}

\subsection{Setup and Data Set}
\label{sec:Setup}
A schematic diagram of the HPGe detector and its surrounding NaI(Tl) anti-Compton (AC) detector system is
presented in Figure 7 of Ref.~\cite{TEXONO:2006xds}. The shielding configuration, including the cosmic-ray (CR) veto scintillators and the data acquisition systems are illustrated in Figures 6 and 8, respectively, of ~Ref.~\cite{TEXONO:2006xds}. The HPGe detector has a fiducial target mass of 1.06 kg and;  reactor-ON and reactor-OFF exposures are 278.91 and 43.60 days, respectively.

\begin{figure}[h]
\includegraphics[width=9cm]{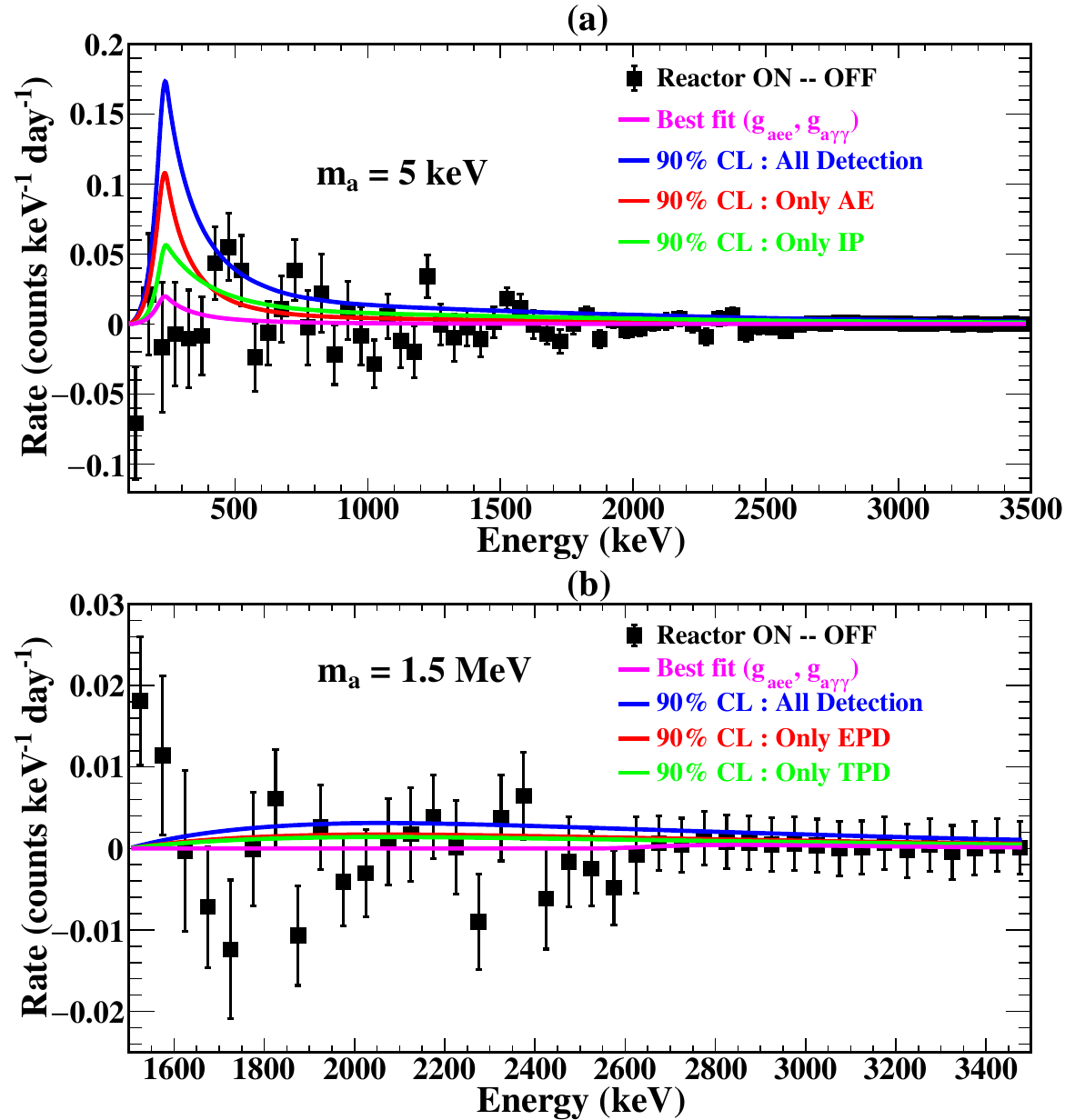}
\caption{Measured reactor ON--OFF residual spectrum (black data points) compared with expected ALP signal distributions for $m_a$ at (a) 5$\text{ keV}$ and (b) $1.5\text{ MeV}$. The corresponding $\gagg$ and $g_{aee}$ coupling constants are taken from the parameters denoted by green stars of Figures~\ref{fig:Global5KeV} and \ref{fig:Global1500KeV}, respectively. The best-fit and  90\% CL spectra, as well as contributions from individual detection channels, are shown. }
\label{fig:On_minus_OFF}
\end{figure}

This data set is selected for the present analysis due to its low background levels and an extended energy range spanning up to 
3.5 MeV$_{ee}$.  Following the event classification convention adopted in Ref.~\cite{TEXONO:2006xds}, HPGe-detector triggered events
are categorized according to their coincidence with the $AC^{\pm}~\otimes~CR^{\pm}$, where the superscripts + and - denote coincidence and anti-coincidence with the AC and CR veto detectors,
respectively~\cite{TEXONO:2013hrh, TEXONO:2018nir}. Events classified as $AC^{-}~\otimes~CR^{-}$ are 
uncorrelated with both active veto systems and are therefore identified as candidate events for neutrino interactions, ALPs, or other exotic processes. Figure~\ref{fig:DataHist} presents the measured HPGe spectra for the reactor ON and OFF data sets. The observed
background peaks originate either from well-known $\gamma$-rays associated with ambient environmental radioactivity or X-rays produced by cosmogenic activation.

\subsection{Analysis and Results}
\label{sec:Result}
The comparison between reactor ON/OFF data provides a robust differential probe for reactor-induced
rare particle searches by effectively suppressing steady-state backgrounds. The reactor-produced ALPs discussed in Section~\ref{sec:formulation}
would manifest themselves as a statistically significant excess in the reactor ON--OFF 
``residual'' energy spectrum as shown in Fig.~\ref{fig:On_minus_OFF}. A minimum-$\chi^{2}$ analysis is performed to determine the best-fit values of the coupling constants $\gagg$ and $\gaee$. A $\pm$10\% systematic uncertainty is assigned to the predicted ALP flux to account for its dependence on the reactor fuel composition, which affects the prompt $\gamma$-ray emission. No statistically significant excess is observed from the reactor ON--OFF residual spectra for any combination of ALP mass and coupling parameters.  Therefore, 90\% confidence level (CL) exclusion limits are derived using the unified approach of Ref.~\cite{FC_PRD1998}. The best-fit ALP signals under the assumption of simultaneously non-zero
$\gagg$ and $\gaee$ couplings, along with their corresponding 90\% CL upper signal rates, are superimposed on the residual energy spectrum in Fig.~\ref{fig:On_minus_OFF} for two representative ALP masses of 5~keV and 1.5~MeV. The values of coupling
constants $\gagg$ and $\gaee$ are adopted from the global exclusion plots, indicated by the green star in Figs.~\ref{fig:Global5KeV} and \ref{fig:Global1500KeV}, respectively. To illustrate the relative contributions of the dominant detection channels for
each representative ALP mass,  the 90\% CL limit obtained from AE and IP processes are also shown separately.  For each individual process, the values of coupling constants $\gagg$ and $\gaee$ are fixed to the intersection point of their respective global exclusion curves.

\subsubsection{Photophilic Limit ($\gagg > 0$, $\gaee = 0$)}

\begin{figure}[t]
\includegraphics[width=8.5cm]{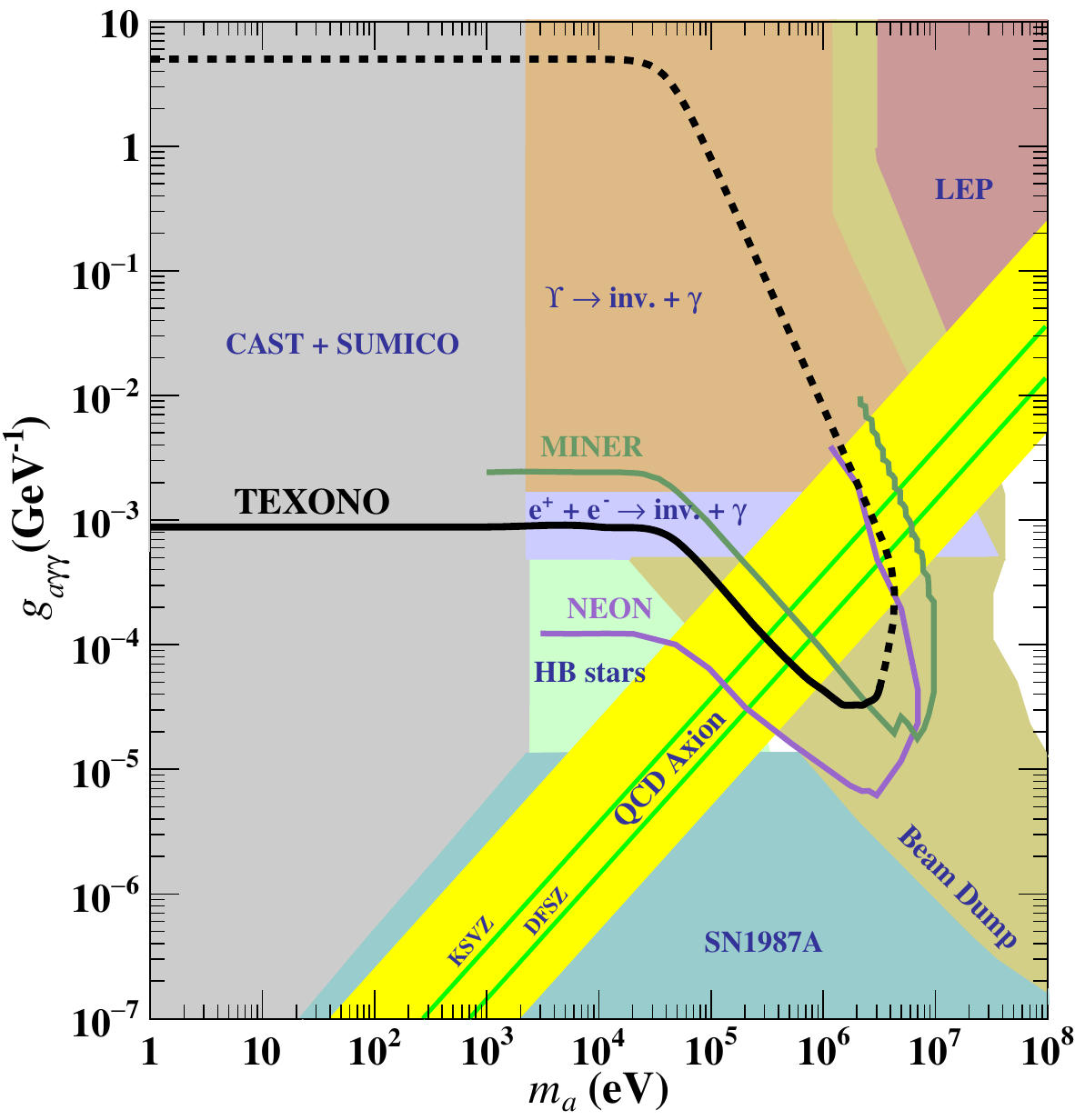}
\caption{Exclusion limits at the 90\% CL in the ($\ma$, $\gagg$) parameter space derived
  from the TEXONO data is shown as the solid black curve (upto 3.2 MeV);  the dotted black line indicates the upper experimental sensitivity limit
  derived by using the ALP survival probability along the baseline. The shaded regions indicate existing cosmological and astrophysical
  constraints adapted from Refs.~\cite{JHEP_2021, Jaeckel:2015jla, Carenza:2020zil, Lucente:2020whw}. Direct laboratory constraints
  from the beam-dump~\cite{Blumlein:1991xh}, MINER~\cite{Mirzakhani:2025bqz}, and NEON~\cite{NEON:2024kwv} experiments are also shown for comparison. The parameter space predicted by the KSVZ QCD axion model~\cite{DILUZIO20201} is also shown in the yellow band included for completeness.}
\label{fig:Limitgagg}
\end{figure}

The 90\% CL exclusion limit for $\gagg$ versus $\ma$ at $\gaee = 0$ is depicted in Figure~\ref{fig:Limitgagg}. The explored mass range for reactor ALPs is $\ma~\le~$3 MeV, which corresponds to the dynamic range of the HPGe measurements. The sensitivity is primarily driven by the IP process at $m_a< 20$ keV/c$^2$,  while TPD becomes dominant at higher $\ma$. At the ultra-relativistic limit, $\Ea >> \ma$, mass-dependent terms become negligible in IP cross section. Therefore, the upper bounds on $\gagg$ becomes mass-independent and manifest as a horizontal contour for $\ma < 10$~ keV/$c^{2}$. The most stringent bound obtained from this analysis is $\gagg < 3.27 \times 10^{-5}$ GeV$^{-1}$ at $\ma = 1.75$ MeV/$c^2$ due to TPD channel. Beyond this mass, the sensitivity deteriorates as the peak of the differential spectrum shifts outside energy range of the measurements. The solid black curve up to 3.2 MeV represents the constraint from this work. The black dotted line indicates the upper sensitivity reach of the experiment, derived by accounting for the survival probability of ALPs during their propagation from the source to the detector.

\begin{figure}[t]
\includegraphics[width=8.5cm]{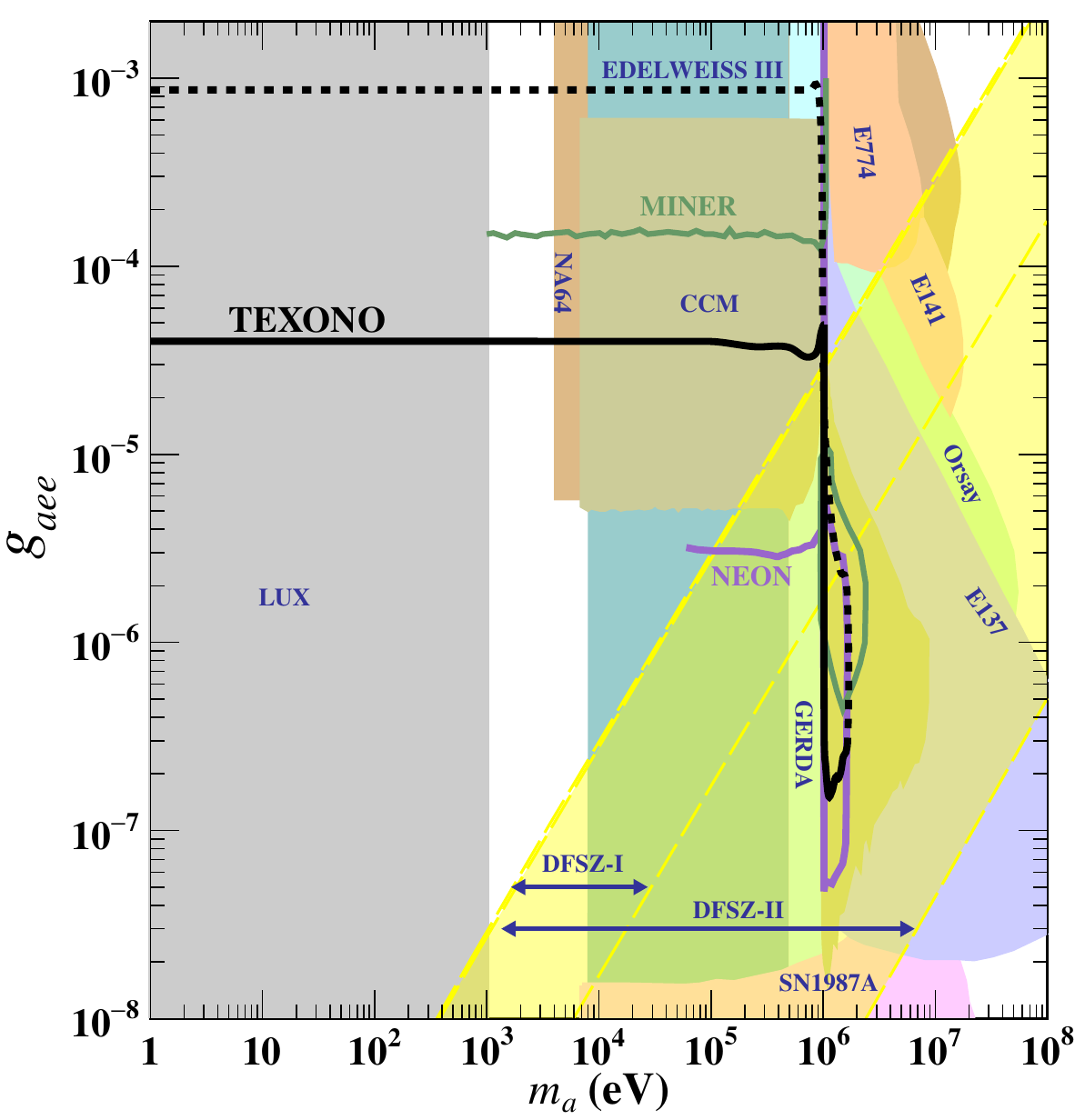}
\caption{Exclusion limits at the 90\% CL in the ($\ma$, $\gaee$) parameter space derived from the TEXONO data are shown as a solid black line; the upper sensitivity reach of the experiment is represented by the dotted black curve. 
  At lower axion masses, the parameter space is constrained by direct solar axion searches from LUX~\cite{LUX:2017glr},  together with dark matter ALP searches from EDELWEISS-III~\cite{EDELWEISS:2018tde} and GERDA~\cite{GERDA:2020emj}. At higher mass scales, limits from beam-dump experiments~\cite{Bechis:1979kp, Riordan:1987aw, Bross:1989mp, NA64:2021aiq, CCM:2021jmk} and SN1987A~\cite{Carenza:2021pcm} provide the most stringent laboratory constraints. The reactor ALP constraints from MINER~\cite{Mirzakhani:2025bqz} and NEON~\cite{NEON:2024kwv} are also shown for comparison. The yellow band depicts the parameter space predicted by the DFSZ-I and DFSZ-II QCD axion models from Ref.~\cite{DILUZIO20201} is also included for completeness.}
\label{fig:Limitgaee}
\end{figure}

Existing laboratory constraints together with cosmological/astrophysical bounds from Refs.~\cite{JHEP_2021, Jaeckel:2015jla, Carenza:2020zil, Lucente:2020whw} are superimposed for comparison. The present results significantly
improve upon the existing experimental limits on the ALP-photon coupling, $\gagg$, previously obtained from reactor-based experiment MINER~\cite{Mirzakhani:2025bqz} and searches for radiative Upsilon decays ($\Upsilon \rightarrow \gamma + \text{invisible}$)~\cite{Masso:1995tw}.
The derived exclusion limits probe and exclude a portion of the parameter space predicted by the KSVZ QCD axion model~\cite{DILUZIO20201}
for axion masses of a few hundred keV/$c^2$. The exclusion limits presented in this work are less stringent than those reported by the reactor experiment NEON~\cite{NEON:2024kwv}, primarily owing to its substantially larger exposure. 

\subsubsection{Photophobic Limit ($\gagg = 0$, $\gaee > 0$)}

The 90\% CL exclusion limit for $\gaee$ as a function of the ALP mass $\ma$ at $\gagg = 0$,  are presented in Fig.~\ref{fig:Limitgaee}. 
For $\ma < 2\me$, the experimental
sensitivity is  mostly $m_{a}$ independent at $m_{a}<<E_{a}$, and is dominated by AE absorption and IC scattering. Above the (2$m_{e}$) threshold, the EPD channel becomes kinematically allow and
dominates the sensitivity. In the present analysis, the exclusion limit on $g_{aee}$ reaches the maximal sensitivity at  $1.61 \times 10^{-7}$ at $1.2\text{ MeV}/c^2$. The sensitivity to $g_{aee}$ is restricted to ALP masses $m_a < 1.75\text{ MeV}/c^2$, corresponding to the maximum accessible energy range of the HPGe measurements. At this mass threshold, the minimum ALP energy ($E_a^{\text{min}}$) produced via the Compton-like process is $3.3\text{ MeV}$, which lies beyond the measured energy window.
Similar to the $\gagg$ exclusion limit in Figure~\ref{fig:Limitgagg}, the upper boundaries of the excluded parameter space are determined
by the finite survival probability of ALPs, which may decay in flight before reaching the detector.

Constraints from representative  benchmark experiments are superimposed for comparison. The LUX~\cite{LUX:2017glr} constraints solar ALPs; while EDELWEISS III~\cite{EDELWEISS:2018tde} and GERDA~\cite{GERDA:2020emj} derive limits under the assumption that ALPs constitute the dark matter. 
Beam-dump constraints are provided by the Orsay~\cite{Bechis:1979kp}, E141~\cite{Riordan:1987aw}, E774~\cite{Bross:1989mp}, NA64~\cite{NA64:2021aiq}, and CCM120~\cite{CCM:2021jmk} experiments. In the mass range $m_a \sim 1 - 200$ MeV/$c^{2}$, the most stringent constraint is provided by Supernova SN1987A~\cite{Carenza:2021pcm}. The limits derived in this work improve upon those the results reported by MINER~\cite{Mirzakhani:2025bqz} for $\ma < 2\me$, but remain less stringent than the corresponding limits from NEON~\cite{NEON:2024kwv}. The predictions for QCD axions~\cite{Kim:1979if, Shifman:1979if} are shown as a yellow band. The present analysis probes only a small region of the parameter space favored by the KSVZ model.

\subsubsection{Case of Active $\gagg$ and $\gaee$}

\begin{figure}[t]
\includegraphics[width=9cm]{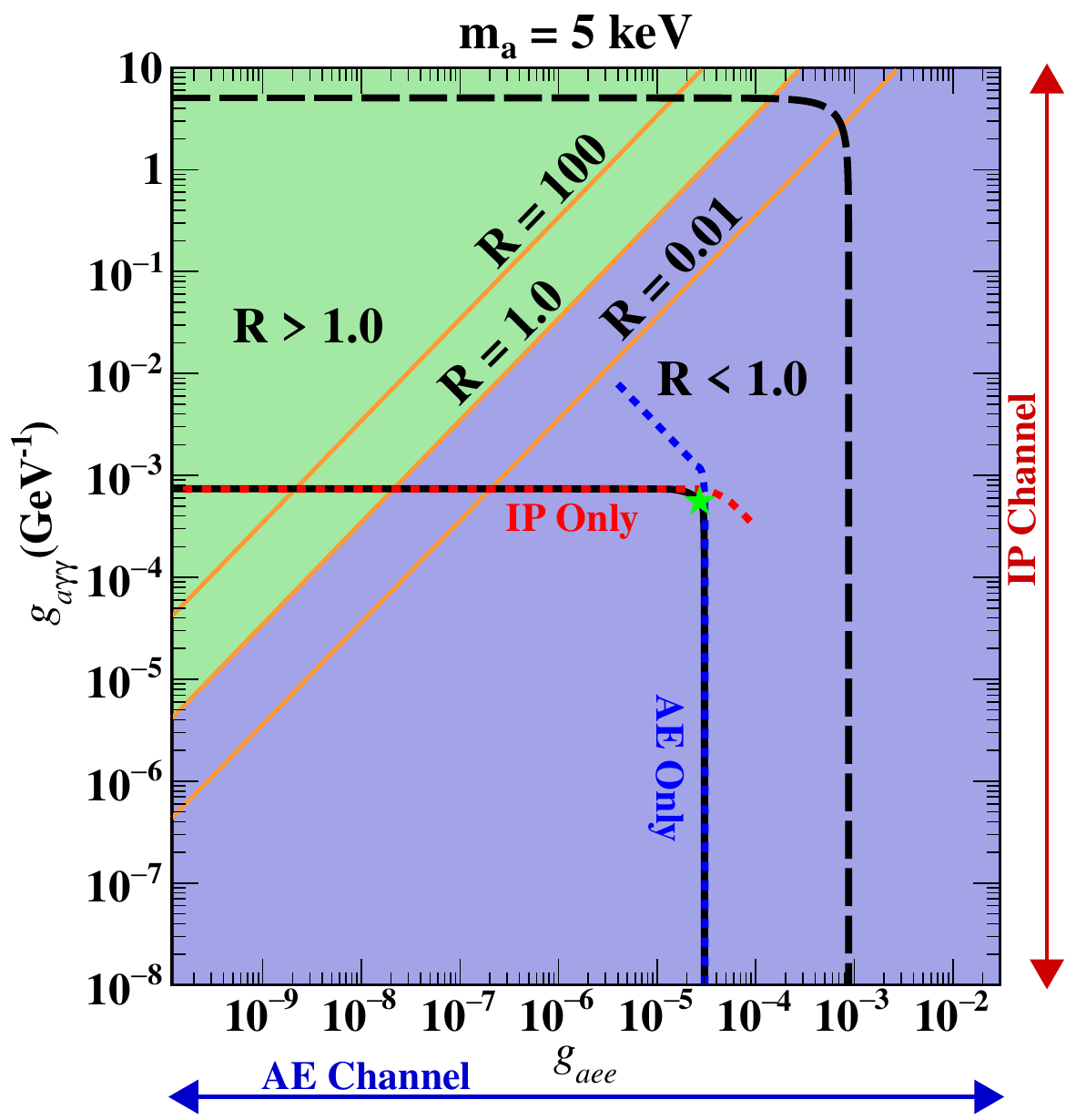}
\caption{The 90\% CL exclusion contour from a two-parameter ($\gaee$, $\gagg$) fit to the reactor ON--OFF data
  is shown as the black solid curve for $m_{a}$ = 5 keV, incorporating all ALP flux components and both dominant detection channels
  (IP and AE). For comparison, the limits that include all flux components but only individual detection channels are also displayed: AE absorption only (blue dotted line) and IP scattering only (red dotted line). Regions labeled $R$ and their corresponding contours indicate ALP production rates across different ratios of $\gagg$ and $\gaee$. The green star indicates the benchmark paramter value for spectrum depicted in Fig.~\ref{fig:On_minus_OFF}(a). The dotted black line indicates the upper experimental sensitivity limit.}
\label{fig:Global5KeV}
\end{figure}
\begin{figure}[t]
\includegraphics[width=9cm]{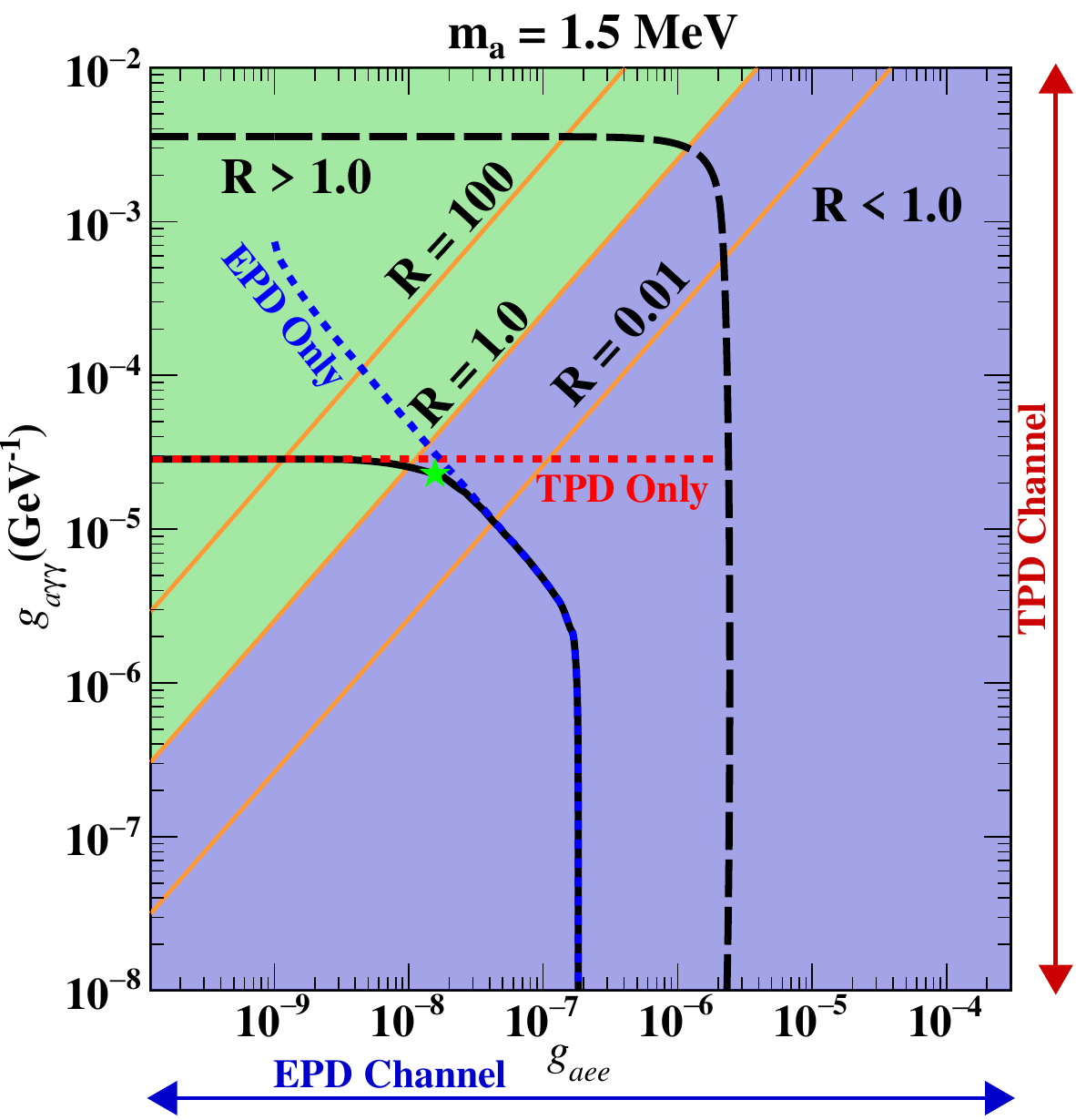}
\caption{The 90\% CL exclusion contour from a two-parameter ($\gaee$, $\gagg$) fit to the reactor ON--OFF data
  is shown as the black solid curve for $\ma$ = 1.5 MeV, incorporating all ALPs flux components and both dominant detection channels
  (TPD and EPD). For comparison, the limits that include all flux components but only individual detection channels are also displayed: EPD only (blue dotted line) and TPD only (red dotted line). The R regions and contours describe the ALP-production from the ratio of $\gagg$ to $\gaee$. The green star corresponds to the spectrum of Fig.~\ref{fig:On_minus_OFF}(b). The dotted black line denotes the upper bound of the experimental sensitivity.}
\label{fig:Global1500KeV}
\end{figure}

Figures~\ref{fig:Global5KeV} and~\ref{fig:Global1500KeV} present the two-dimensional 90\% CL exclusion regions in the ($\gaee$, $\gagg$) parameter space for fixed ALP masses of $\ma$ = 5 keV and 1.5 MeV, respectively. The derived upper limits and expected sensitivity boundaries are represented by the solid and dashed black contours, respectively, with the region enclosed between them being experimentally probed and excluded.
The ALP signal is suppressed by material shielding and a short decay length at large values of $\gagg$ and $\gaee$,  defining the upper sensitivity boundaries.

The ratio of the ALP flux contribution from the Primakoff ($\gagg$) process to that from the Compton-like ($\gaee$) process is denoted by $R$.
The light blue and light green shaded regions indicated parameter space dominant ALP production mechanism is $via$ $\gaee$ (R < 1) and/or $\gagg$ (R > 1), respectively.

As shown in Figure~\ref{fig:Global5KeV}, for $m_a = 5\text{ keV}$, the dominant detection channels associated with the $g_{a\gamma\gamma}$ and $g_{aee}$ couplings are IP scattering and AE effect, respectively. These conditions are generally satisfied for light ALPs with $m_a < 20\text{ keV}$. In the higher-mass regime, distinct behaviors emerge for $\gagg$ coupling, the TPD channel becomes dominant when $m_a > 20\text{ keV}$, whereas for the $\gaee$ coupling, the EPD channel becomes dominates for $m_a > 1.023\text{ MeV}$. The resulting 90\% CL exclusion contour for $m_a = 1.5\text{ MeV}$ is presented in Fig.~\ref{fig:Global1500KeV}.

\section{Conclusion \label{sec:summary}}
The direct detection of ALPs is of considerable interest because they are well-motivated dark matter candidates.
In addition to cavity and beam-dump based searches, nuclear reactor experiments provide a powerful and complementary approach to the direct detection of ALPs. In this work, we present a comprehensive
analysis of data collected with a 1.06 kg HPGe detector to constrain both the $\gagg$ and $\gaee$
couplings over the ALP mass range from 1 eV to 3 MeV. Despite the relatively modest detector mass and exposure,
the present analysis improves the existing constraints on both couplings
in the MeV mass region compared with those obtained by significantly large experiments. In particular, for the ALP-electron
coupling $g_{aee}$ in the sub-MeV mass regime ($m_a \lesssim 1$ MeV), our results probe
previously unexplored parameter space that had been accessible only through dark matter direct-detection experiments.
These results demonstrate the capability of compact, low-background reactor-based experiments to complement,
and in certain region, surpass the sensitivity of much larger-scale facilities and astrophysical probes
in the search for light ALPs.

\begin{acknowledgments}
  This work is supported by the Investigator Award No. AS-IA-106-M02 and
  Thematic Project No. AS-TP-112-M01 from the Academia Sinica, Taiwan, and
  Contract No. 106-2923-M-001-006-MY5, No. 107-2119-M-001-028-MY3,
  No. 110-2112-M-001-029-MY3, and No. 113-2112-M-001-053-MY3
  from the National Science and Technology Council, Taiwan (H. T. W.);
  Contract F.30-584/2021 (BSR), UGC-BSR Research Start Up Grant, India;
  the PURSE grant  and FIST program of Department of Science \& Technology,
  India (L. S. and V. S). Greeshma~C. and D.~K.~Mishra gratefully
  acknowledge the financial support provided by Taiwan International Graduate Program (TIGP-X) fellowship of
  Academia Sinica. 

\end{acknowledgments}

\bibliography{recALP}

\end{document}